\documentclass[10pt]{article}
\usepackage{latexsym,amsmath,amssymb,amsfonts,enumerate}
\usepackage[mathscr]{eucal}
\usepackage{graphicx}

\newcommand{\gra}{\alpha}
\newcommand{\grb}{\beta}
\newcommand{\grg}{\gamma}
\newcommand{\grd}{\delta}

\newcommand{\grl}{\lambda}

\newcommand{\grs}{\sigma}

\newcommand{\grt}{\tau}

\newcommand{\grf}{\varphi}

\newcommand{\real}{\mathbb{R}}

\renewcommand{\cal}{\mathscr}
\newcommand{\p}{\partial}

\newcommand{\dal}{\square}
\newcommand{\Lag}{\cal{L}}

\DeclareMathOperator{\rank}{rank}
\newcommand{\unitdn}[1]{\hat{k}_{#1}}
\newcommand{\unitup}[1]{\hat{k}^{#1}}
\newcommand{\anisotropy}{k}
\newcommand{\opstwo}[3]{ \underset{#1#2}{\mathscr{#3}}}
\newcommand{\opsthree}[4]{ \underset{#1#2#3}{\mathscr{#4}}}
\newcommand{\opg}[3]{ \underset{#1#2\{#3\}}{\mathscr{G}}}

\newcommand{\connfirst}[4]{\overset{(#1)}{\Gamma}{}_{#2#3#4}}
\newcommand{\connsecond}[4]{\overset{(#1)}{\Gamma}{}_{#2#3}^{#4}}

\begin{document}
\title{
\bf \huge{ A Geometrical Anisotropic Model \\ of Space-time
Based  on \\ Finslerian Metric}}
\author{P.C. Stavrinos\\
Department of Mathematics   \\ 
University of Athens, Greece\\ 
pstavrin@cc.uoa.gr
     \and
F.I. Diakogiannis   \\ 
Tichis 9, 19010 \\ 
Lagonisi, Greece}
\date{}
\maketitle
\begin{abstract}
  In this work we study  an anisotropic model of general relativity
based on the framework of Finsler geometry. The observed anisotropy
of the microwave background radiation \cite{Peebles} is incorporated
in the Finslerian structure of
space-time.
We also examine the electromagnetic (e.m.) field equations in our space.
 As a result a modified wave equation of e.m. waves yields.
\end{abstract}
\section{Introduction}
A Finslerian geometrical structure
of models which can correspond to aniso-tropic structures of regions
of spacetime (radius$\leq 10^8$ light years)
can be introduced. Our work was
motivated by the observed anisotropy of the microwave cosmic
radiation. This anisotropy is of dipole type, i.e. the intensity of
the radiation is maximum at one direction and minimum at the opposite
direction.

It is known that this anisotropy can be explained if we
use Robertson-Walker metric and take into account the movement of our
galaxy with respect to distant galaxies of the universe \cite{Wald}.
A small anisotropy is expected, however, due to the anisotropic distribution of galaxies 
in space \cite{Peebles}.

From the above mentioned results, it is reasonable
 to seek for a Lagrangian which expresses this anisotropy. As such,  we choose:
\begin{equation}
\Lag=\sqrt{a_{ij}y^{i}y^j}+\grf(x)\unitdn{a}y^{a}
\end{equation}
The vector $\unitdn{a}$ expresses the observed anisotropy of the microwave background 
radiation.

In \S \ref{Preliminaries} we give the necessary mathematical formalism, upon which we 
develop our theory.

In \S \ref{main} we develop the geometric anisotropic structure
of space-time based on the tangent bundle. Some physical interpretations are given.

In  \S \ref{electro} we study the changes that are imposed on the electromagnetic field 
as a result of the anisotropic geometry. It is shown that the e.m. field tensor remains 
unchanged in our approach. The wave equation of e.m. waves is modified
($\dal_F A^{i}(x)$) in such a way that it expresses  anisotropy of the electromagnetic 
field, i.e. in the generalized D'Alambertian there exist terms of anisotropy which 
affect the conventional form of the wave equation.

\section{Preliminaries } \label{Preliminaries}
The framework in which we develop our present work is a Finsler tangent bundle. For this 
we consider a smooth $4$-dimensional
pseudoriemannian manifold M, $(TM,\pi , M)$ its tangent bundle and
$\tilde{TM}=TM\setminus\{ 0 \}$, where $0$ means the image of the null cross-section of 
the projection $\pi : TM \rightarrow M$.
We also consider a local system of coordinates $(x^{i}),\, i=0,1,2,3$
and $U $  a chart of $M$. Then the couple   $(x^{i},y^{a})$  is a local system of 
coordinates on $\pi^{-1} (U)$ in $TM$. A coordinate
transformation on the total space $TM$ is given by
\begin{equation} \label{1-1}
\tilde{x}^{i}=\tilde{x}^{i}(x^{0},\ldots,x^3),\quad \det \left\|
\frac{\p \tilde{x}^{i}}{\p x^j}
\right\|\neq 0, \quad \tilde{y}^{a}= \frac{\p
\tilde{x}^{a}}{\p x^b} y^b,
\quad x^{a}=\grd^{a}_{i}x^{i}
\end{equation}

By definition \cite{Miron1} a Finsler metric on M is a function
$F : TM \to \real $ having the properties:
\begin{enumerate}

\item The restriction of $F$ to $\tilde{TM}$ is of the class $C^{\infty}$ and $F$ is 
only continuous on the image of the null cross section in the tangent bundle to $M$.

\item   The restriction of $F$ to $\tilde{TM}$ is positively
homogeneous of degree 1 with respect to $(y^{a})$.
\[
F(x,ky)=kF(x,y),\quad k\in \real^{*}_{+}
\]

\item The quadratic form on $\real^n$ with the coefficients
\begin{equation} \label{metric}
f_{ij}= \frac{1}{2} \frac{\p^2 F^2}{\p y^{i}\p y^{j}}
\end{equation}
defined on $\tilde{TM}$ is non degenerate ($\det(f_{ij})\neq 0$), with 
$\rank(f_{ij})=4$.
\end{enumerate}

As it is known a non linear connection $N$ on $TM$ is a distribution on $TM$, 
supplementary to the vertical distribution $V$ on $TM$ :
\[
 T_{(x,y)}(TM)=N_{(x,y)}\oplus V_{(x,y)}
\]
In our case a non linear connection can be defined by
\begin{equation} \label{Naj}
N^{a}_j= \frac{\p G^{a}}{\p y^j}
\end{equation}
where $G^{a}$ are defined from
\begin{equation} \label{Gl}
\quad G^{a}= \frac{1}{4} f^{aj}
\left(
\frac{\p^2 \Lag}{\p y^j\p x^k}y^k-\p_j\Lag
\right)
\end{equation}
and the relation
\begin{equation} \label{Gl2}
\frac{dy^{a}}{ds}+2G^a(x,y)=0
\end{equation}
 yields from the  \mbox{Euler-Lagrange} equations:
\begin{equation} \label{EulerLagrange}
 \frac{d}{ds}\left(
 \frac{\p \Lag}{\p y^{a}}
 \right)- \frac{\p \Lag}{\p x^{a}}=0
\end{equation}

The transformation rule of the non-linear connection coefficients
is
\begin{equation} \label{transnonlinear}
\tilde{N}^{a}_{i}= \frac{\p \tilde{x}^{a}}{\p x^{b}}
\frac{\p x^{j}}{\p \tilde{x}^{i}}N^b_j(x,y)+
\frac{\p \tilde{x}^{a}}{\p x^{h}}
\frac{\p^2 x^h}{\p \tilde{x}^{i} \p \tilde{x}^{b}}y^b
\end{equation}
also
\begin{align*}
\frac{\grd}{\grd \tilde{x}^{i}}
&= \frac{\p x^{j}}{\p \tilde{x}^{i}} \frac{\grd}{\grd x^j}
& \frac{\p}{\p \tilde{y}^{a}}&= \frac{\p x^b}{\p\tilde{x}^{a}}
\frac{\p}{\p y^{b}}\\
d\tilde{x}^{i}&=
\frac{\p \tilde{x}^i}{\p x^j}dx^j
&\grd \tilde{y}^{a}&=
\frac{\p \tilde{x}^{a}}{\p x^b}\grd y^{b}
\end{align*}

A local basis
of $T_{(x,y)}(TM)$, $(\grd_i,\dot{\p}_a)$
adapted to the horizontal distribution $N$ is
\begin{equation} \label{delta}
\grd_i=\p_i-N^a_i(x,y)\dot{\p}_a ,\quad \text{where}\quad
\p_i= \frac{\p}{\p x^{i}},\quad
\dot{\p}_a= \frac{\p}{\p y^{a}}
\end{equation}
where $N^{a}_i(x,y)$ are the coefficients of the non-linear Cartan connection N as we 
mentioned above.\newline
The dual local basis is
\begin{equation*}
\{
d^{i}=dx^{i},\, \grd^{a}=\grd y^{a}=dy^{a}+N^{a}_jdx^j
\}_{i,a=\overline{0,3}}\equiv
\{
\grd^{\grb}
\}_{\grb=\overline{0,7}}
\end{equation*}

A $d$-connection on tangent bundle $TM$ is a linear connection on $TM$ which preserves 
by parallelism the horizontal distribution
$N$ and the vertical distribution $V$ on $TM$.

Generally an $h$-$v$ metric on the tangent bundle
$(TM,\pi , M)$ is given by
\begin{equation} \label{G}
G=f_{ij}(x,y)dx^{i}\otimes dx^{j}+ h_{ab} \grd y^{a}\otimes
\grd y^{b}
\end{equation}

We consider a metrical d-connection
 $C\Gamma =(N^{a}_j,L^{i}_{jk},C^{i}_{jk})$
 with the property
\begin{align} \label{*}
f_{ij|k}&=\grd_k f_{ij}- L^h_{ik}f_{hj}-L^h_{jk}f_{ih}=
0\\
f_{ij}|_k&=\dot{\p}_k f_{ij}- C^h_{ik}f_{hj}-C^h_{jk}f_{ih}=0
\label{**}
\end{align}
where
\begin{align} \label{connectionL}
L^{i}_{jk}&= \frac{1}{2} f^{ir}\left( \grd_j f_{rk}+\grd_k
f_{jr}-\grd_r f_{jk} \right)\\ \label{connectionC}
 C^{i}_{jk}&= \frac{1}{2}f^{ir}
 (
\dot{\p}_j f_{rk}+\dot{\p}_k f_{jr}-\dot{\p}_r f_{jk}
)
\end{align}
The coordinate transformation of the objects $L^{i}_{jk}$ and
$C^{i}_{jk}$ is:
\begin{align} \label{transLijk}
\tilde{L}^{i}_{jk}&= \frac{\p \tilde{x}^{i}}{\p x^h}
\frac{\p x^l}{\p\tilde{x}^j} \frac{\p x^r}{\p \tilde{x}^k}
L^h_{lr}(x,y)+ \frac{\p \tilde{x}^{i}}{\p x^r}
\frac{\p^2 x^r}{\p \tilde{x}^j\p \tilde{x}^k}\\
\tilde{C}^{i}_{jk}&=
\frac{\p \tilde{x}^{i}}{\p x^h}
\frac{\p x^l}{\p\tilde{x}^j} \frac{\p x^r}{\p \tilde{x}^k}C^{h}_{lr}(x,y)
\end{align}

The Cartan torsion coefficients $C_{ijk}$ are given by
\begin{equation} \label{Cartan}
C_{ijk}= \frac{1}{2}\dot{\p}_k f_{ij}
\end{equation}
 while the Christoffel
symbols  of the first and second kind  for the metric $f_{ij}$ are
respectively:
\begin{align} \label{gamma1}
\grg_{ijk}&= \frac{1}{2} \left( \frac{\p f_{kj}}{\p x^{i}}+
\frac{\p f_{ik}}{\p x^{j}}-\frac{\p f_{ij}}{\p x^{k}}
\right)\\
 \label{gamma2}
\grg_{ij}^l&= \frac{1}{2}f^{lk} \left( \frac{\p f_{kj}}{\p x^{i}}+
\frac{\p f_{ik}}{\p x^{j}}-\frac{\p f_{ij}}{\p x^{k}}
\right)
\end{align}

The torsions and curvatures which we use are given by \cite{Miron1,Miron2}:
\begin{align} \label{torsion1}
T^{i}_{kj}&=0 &S^{i}_{kj}&=0
&R^{i}_{jk}&=\grd_k
N^{i}_j- \grd_j N^{i}_k
\\\label{torsion2}
P^{i}_{jk}&=\dot{\p}_{k}N^{i}_j-L^{i}_{kj}
&P^{i}_{jk}&=f^{im}P_{mjk}
&P_{ijk}&=C_{ijk|l}\,y^{l}
\end{align}
\begin{align} \label{Rijkl}
R^{i}_{jkl}&=\grd_l L^{i}_{jk} \grd_k L^{i}_{jl}+
L^{h}_{jk}L^{i}_{hl}-L^{h}_{jl}L^{i}_{hk}+ C^{i}_{jc}R^{c}_{kl}\\
\label{Sijkldn} S_{jikh}&=C_{iks}C^{s}_{jh}-C_{ihs}C^{s}_{jk}\\
\label{Pijkldn}
P_{ihkj}&=C_{ijk|h}-C_{hjk|i}+C^{r}_{hj}C_{rik|l}\,y^l-
C^{r}_{ij}C_{rkh|l}\,y^l\\\label{Sijklup}
S^{l}_{ikh}&=f^{lj}S_{jikh}\\\label{Pijklup}
 P^{l}_{ikh}&=f^{lj}P_{jikh}
\end{align}
\section{The Geometrical Structure of the anisotropic model
(based on $TM$)}\label{main}

In the following, the lowering and raising of the indices of the
objects $\unitdn{a},y^{a}$ and all related Riemannian
tensors will be performed with the metric $a_{ij}$. For the
related Finslerian tensors we shall use the Finsler metric
$f_{ij}$.

The Lagrangian which gives the equation of geodesics in the case of
(pseudo)-Riemannian space-time is given by:
\begin{equation} \label{LR}
L=\sqrt{a_{ij}y^{i}y^{j}}, \quad y^{i}= \frac{dx^{i}}{ds}
\end{equation}
or, equivalently, we may write  for the line element:
\begin{equation} \label{dsR}
ds_R=\sqrt{a_{ij}dx^{i}dx^j}
\end{equation}
where $a_{ij}$ is the Riemannian metric with signature $(-,+,+,+)$.
Because of the observed anisotropy, we must insert an additional term to the Riemannian 
line element \eqref{dsR}. This term must fulfill  the
following requirements:
\begin{enumerate}[(a)]

\item It must give absolute maximum contribution for direction of
movement parallel to the anisotropy axis.

\item It must give zero contribution for movement in direction perpendicular to the 
anisotropy axis, i.e. the new line element must coincide with the Riemannian one for 
direction vertical to the anisotropy axis.

\item It must not be symmetric with respect to replacement
$y^{a}\to - y^{a}$. This requirement is necessary in order to express
the anisotropy of dipole type of the Microwave   Background Radiation
(MBR). We need to have maximum (positive) contribution for direction that coincides with 
the direction of the anisotropy axis, and minimum (negative) contribution for the 
opposite direction.
\end{enumerate}

We see that a term which satisfies the above conditions is
$\anisotropy_a(x)y^{a}$, where $\anisotropy_a(x)$ expresses this
anisotropy axis. For constant direction of $\anisotropy_a(x)$ we may consider 
$\anisotropy_a(x)=\grf(x)\unitdn{a}$, where $\unitdn{a}$ is the unit vector in the 
direction $\anisotropy_a(x)$. Then $\grf(x)$
plays the role of ``length'' of the vector $\anisotropy_a(x)$, $\grf(x) \in \mathbb{R}$. 
Hence, we have the Lagrangian
\begin{equation} \label{lagrangian2}
\Lag=\sqrt{a_{ij}y^{i}y^j}+\grf(x) \unitdn{a} y^{a}
\end{equation}
From \eqref{lagrangian2} we define the Finsler metric function
$F(x,y)=\Lag$. Setting $y^{a}=dx^{a}$ we have
\begin{equation} \label{dsF}
ds_F=\sqrt{a_{ij}dx^{i}dx^j}+\grf(x) \unitdn{a} dx^{a}
\end{equation}
$ds_F$ is the Finslerian line element and  $ds_R$ is the Riemannian
one. We see that the Finslerian line element is generated by an
additional increment to the Riemannian one due to the
anisotropy axis. Now
\begin{equation} \label{ds2F}
ds^2_F=a_{ij}dx^{i}dx^j+2\grf(x)\unitdn{a}dx^{a}\sqrt{a_{ij}
dx^{i}dx^j}
+\grf^2(x)\unitdn{a}dx^{a}\unitdn{b}dx^{b}
\end{equation}
 In order for the
Finslerian metric to be physically consistent with General Relativity
theory, it must have the same signature with the Riemannian metric
$(-,+,+,+)$. We have
\begin{equation} \label{dsRgamma}
ds_R=c\, d\grt=c\, \grg dt = \grg d(ct)=\grg dx^0
\end{equation}
where $\grg=\sqrt{1-(v/c)^2}$ and $v$: 3-velocity in Riemannian space-time.
From relations \eqref{dsRgamma},\eqref{ds2F} we obtain:
\begin{align} \notag
ds^2_F&=a_{00}dx^0dx^0+
 2a_{0\gra}dx^{0}dx^{\gra}+a_{\gra\grb}dx^{\gra}dx^{\grb}+
2\grf(x)\unitdn{0}dx^{0}ds_R
\\ \notag
&+2\grf(x)\unitdn{\gra}dx^{\gra}ds_R
+\grf^2(x)\unitdn{0}\unitdn{0}dx^{0}
dx^{0}
+2\grf^2(x)\unitdn{0}\unitdn{\gra}dx^{0}
dx^{\gra}
\\ \notag
&+\grf^2(x)\unitdn{\gra}dx^{\gra}
\unitdn{\grb}dx^{\grb}
\end{align}
or
\begin{align}
ds^2_F&=\left(
a_{00}+2\grg\grf(x)\unitdn{0}+\grf^2\unitdn{0}\unitdn{0}
\right)dx^{0}dx^{0}+  \notag\\ \notag
&+\left(a_{\gra\grb}+
\grf^2(x)\unitdn{\gra}
\unitdn{\grb}
\right)dx^{\gra}dx^{\grb}
+2\grg\grf(x)\unitdn{\gra}dx^{\gra}dx^{0}
\\
&+2a_{0\gra}dx^{0}dx^{\gra}+
2\grf^2(x)\unitdn{0}\unitdn{\gra}dx^{0}
dx^{\gra}
\label{f00}
\end{align}
where $\gra,\,\grb=1,2,3$.
From relation \eqref{f00} it is evident that we must have
\begin{align} \label{binom}
(\anisotropy_{0}(x))^2+2\grg\anisotropy_{0}(x)+a_{00}&< 0\\
\grd^{\gra\grb}\left(a_{\gra\grb}+
\anisotropy_{\gra}(x)
\anisotropy_{\grb}(x)\right)
&>0 \label{ab}
\end{align}
for the signature to be preserved,
where we have written  $\grf(x)\unitdn{i}=\anisotropy_{i}(x)$. Relation \eqref{binom} 
admits negative values for
\begin{equation} \label{interval}
-\grg-\sqrt{\grg^2-a_{00}}<\anisotropy_{0}(x)<-\grg
+\sqrt{\grg^2-a_{00}}
\end{equation}
while from \eqref{ab} yields:
\begin{equation} \label{intervalab}
\left(k_{\gra}(x)\right)^2>-a_{\gra\gra}
\end{equation}
which is true for any $\anisotropy_{a}(x)$ since $a_{\gra\gra}>0$.

Then for any physically acceptable vector, its $0$ component $\anisotropy_{0}(x)$ must 
lie
in the interval \eqref{interval}. Relation \eqref{interval} is a restriction upon the 
anisotropy of space-time, i.e. the anisotropy vector can not take arbitrary values.
\\

The equation of geodesics is given by:
\begin{equation} \label{geodesics}
\frac{d^2 x^{l}}{ds^2}+ \connsecond{a}{i}{j}{l}y^{i}y^j+
\grs a^{lm}(\p_j\grf\unitdn{m}
-\p_m\grf\unitdn{j})y^j=0
\end{equation}
We observe that in the equation of geodesics we have an additional term, namely $\grs 
a^{lm}(\p_j(\grf\unitdn{m})
-\p_m(\grf\unitdn{j}))y^j$ which expresses rotation of the aniso-tropy axis.

Now for the case of electromagnetic waves we must modify relation
\eqref{geodesics}. This is because the world line of an e.m. wave is
null. In geometrical optics the direction of propagation of a light
ray  is determined by the wave vector tangent to the ray. Let 
$\overset{w}{k}{}^l=dx^l/d\grl$ be the four-dimensional wave vector, where $\grl$ is 
some parameter varying along the ray. We have:
\begin{equation} \label{light}
\frac{d \overset{w}{k}{}^l}{d\grl}+ 
\connsecond{a}{i}{j}{l}\overset{w}{k}{}^i\overset{w}{k}{}^j+
\grs a^{lm}(\p_j\grf\unitdn{m}
-\p_m\grf\unitdn{j})\overset{w}{k}{}^j=0
\end{equation}

One possible explanation of the anisotropy axis could be that it
expresses the resultant of the spin densities of the angular momenta
of galaxies in a restricted region of space ($\anisotropy_{a}(x)$ spacelike).
It is known that the mass is anisotropically distributed for regions
of space with radius$\leq 10^8$ light years \cite{Gravitation}.
Then an important kind of anisotropy might result from the ordering of the angular 
momenta of galaxies. As we move to greater distances
(radius$\geq 10^8$ l.y.) the resultant of the spin densities is approximately zero, as 
it is expected for an isotropic universe.
\begin{equation} \label{sumspindensity}
\anisotropy_{a}(x)=\sum_i\overset{(i)}{\anisotropy}{}_{a}(x)\approx 0
\end{equation}
where $\overset{(i)}{\anisotropy}{}_{a}(x)$ is the spin density tensor of each rotating 
mass distribution.

 The spin  is defined
through the spin density tensor \cite{kopz} from the relation
\begin{equation}
S_{ab}= \frac{\sqrt{-g}}{4\pi} \epsilon_{abc}\anisotropy^{c}(x)
\end{equation}

In the case that  $\grf(x)\unitdn{a}$ expresses spin density,
the function $\grf(x )$ is related to mass density (angular momenta
depends upon angular velocity and mass distribution).

From equation
\eqref{geodesics} we see that for small variation of the
resultant of the spin densities vector, the deviation from the
Riemannian geodesics is very small, if not negligible.

From the equation of geodesics \eqref{geodesics} we obtain for
movement $y^i$ perpendicular to $k^i$:
\begin{equation} \label{vertical}
\frac{d^2 x^{a}}{ds^2}+ \connsecond{a}{i}{j}{l}y^{i}y^j+
\grs a^{lm}\p_j\grf\unitdn{m}y^j=0
\end{equation}
From \eqref{vertical} it is evident that although the
contribution to the $ds_R$ line element is zero for
$y^i$ vertical to $k^i$, the equation of geodesics
is different from the Riemannian case. In the case, however, where
$\p_i\grf(x)$is parallel to  $ \unitdn{i}$, i.e. the increment of anisotropy
takes place only along the anisotropy axis, then the equation of geodesics is identical 
with the geodesics of the Riemannian space-time.



Using the notation $\grb=\hat{k}_a y^{a}$, $\grs=\sqrt{a_{ij}y^{i}y^{j}}$, we calculate 
the metric tensor from  \eqref{metric}:
\begin{equation} \label{metricfull}
f_{ij}= \frac{F}{\grs}a_{ij}+ \frac{\grf(x)}{2\grs} \, 
\underset{ij}{\mathscr{S}}\left(y_{i}\hat{k}_j
\right)- \frac{\grb \grf(x)}{\grs^3} y_i y_j + \grf^2(x)\hat{k}_i \unitdn{j}
\end{equation}
where $\underset{ij}{\mathscr{S}}$ is an operator and denotes symmetrization of the 
indices $i,j$, e.g.
\[
\opstwo{i}{j}{S}(A_{ikjl})= \frac{1}{2}(A_{ikjl}+A_{jkil}).
\]
Accordingly we define the antisymmetric operator
\[
\opstwo{i}{j}{A}(M_{ikjl})= \frac{1}{2}(M_{ikjl}-M_{jkil}).
\]

 The inverse metric is
\begin{equation} \label{inverse}
f^{ij}= \frac{\grs}{F} a^{ij}- \frac{\grs\grf}{2F}\,\opstwo{i}{j}{S}
\left(y^{i}\unitup{j}\right)+
\frac{\grf (\grb+m\grs\grf)}{F^3} y^{i}y^j
\end{equation}
as it may be verified by direct calculation, where $m=\unitdn{a}\unitup{a}=0,\pm 1$ 
according whether $\unitdn{a}$ is null, spacelike or timelike (in order to not loose 
generality, we do
not identify $\unitdn{a}$ as spacelike). It must be noted, however, that if $y^{a}$ 
represents the velocity of a particle ($y^{i}$ timelike) then $\unitup{a}$ is bound to 
be spacelike. This follows from the fact that one possible value of $y^{a}\unitdn{a}$ is 
zero.

 The determinant of the
metric is
\begin{equation} \label{determinant}
f=\det (f_{ij})= \left(\frac{F}{\grs}\right)^5\det (a_{ij})
\end{equation}
The Cartan torsion coefficients which are given by \eqref{Cartan}, take the form:
\begin{equation} \label{cartan}
C_{ijl}= \frac{3\grb \grf}{2\grs^5}y_{i}y_{j}y_{l}+
\frac{3\grf}{\grs}\,\opsthree{i}{j}{l}{S}(a_{ij}\unitdn{l})-
\frac{3\grf}{\grs^3}\,\opsthree{i}{j}{l}{S} (y_iy_jy_l)-
\frac{3 \grb \grf}{\grs^3}\, \opsthree{i}{j}{l}{S} a_{ij}y_l
\end{equation}
We observe from \eqref{cartan} that an increment of the anisotropy,
i.e. increment of $\grf$, results in a change in the values of the
components of the Cartan coefficients. This is expected since
the condition
\begin{equation} \label{cartanriemann}
C_{ijk}=0
\end{equation}
is the condition for the Finsler metric to be Riemannian.

The finslerian Christoffel symbols of the first kind are given by
\eqref{gamma1}
\begin{equation} \label{cristoffelfirst}
\grg_{ijl}= \frac{F}{\grs} \connfirst{a}{i}{j}{l}+
\Lambda_{ijl}+M_{ijl}
\end{equation}
where
\begin{equation}
\connfirst{a}{i}{j}{l}= \frac{1}{2} (\p_i a_{lj}+\p_ja_{il}-\p_la_{ij})
\end{equation}
are the Christoffel symbols corresponding to the metric $a_{ij}$.
\begin{equation} \label{Lamda}
\Lambda_{ijl}= \opg{i}{j}{l}\left[\left(
\frac{3\grb\grf}{2\grs^5} y_iy_j- \frac{\grf}{\grs^3}\,\opstwo{i}{j}{S} y_i \unitdn{j}- 
\frac{\grf
\grb}{4\grs^3}a_{ij} \right)\p_la_{ab}y^{a}y^{b}\right]
\end{equation}
and
\begin{equation} \label{Mijk}
M_{ijl}= \opg{i}{j}{l} \left[\left(
\frac{\grb}{2\grs} a_{ij}+ \frac{1}{\grs} \opstwo{i}{j}{S} y_i\unitdn{j}- 
\frac{\grb}{\grs^3} y_i y_j +2\grf \unitdn{i}\unitdn{j} \right)\p_l \grf\right].
\end{equation}
The operator $\opg{i}{j}{l}$ denotes an interchange of the indices in the form this 
interchange appears in the definition of the Christoffel symbols of a metric, e.g.
\begin{align*}
\opg{i}{j}{l}A_{ijl}&=A_{lji}+A_{ilj}-A_{ijl}\\
\opg{i}{j}{l}\p_la_{ij}&=2\connfirst{a}{i}{j}{l}
\end{align*} The
Christoffel symbols of the second kind are found from
\eqref{gamma2}:
\begin{multline} \label{Christoffelsecond}
\grg_{ij}^{l}=\connsecond{a}{i}{j}{l}+\left(
\frac{\grf(\grb+m\grs\grf)}{\grs F^2} y^{a}y^l- \frac{2\grf}{F}
\, \opstwo{a}{l}{S} (y^{a}\unitup{l})
\right)\connfirst{a}{i}{j}{a}+
\frac{\grs}{F} \bigl(
\Lambda_{ij}^{l}+ \\
+M_{ij}^{l}
\bigr)
+(
\Lambda_{ija}+M_{ija})\left(
\frac{\grf(\grb+m\grs\grf)}{F^3}y^{a}y^l-
\frac{2\grs\grf}{F^2}
\opstwo{a}{l}{S}(y^{a}\unitup{l})
\right)
\end{multline}
where $\Lambda^{i}_{jl}=\Lambda_{jlk}a^{ik}$ and
$M^{i}_{jl}=M_{jlk}a^{ik}$.
In relation \eqref{Christoffelsecond} it is seen that, besides the
$\connsecond{a}{j}{k}{i}=0$ terms, the rest express the anisotropic deviation from the
Riemannian Christoffel symbols. When $\grf=0$, i.e. absence of anisotropy, the Finsler
Christoffel symbols coincide with the Riemannian ones. From the above relation, for
$\connsecond{a}{j}{k}{i}=0$ we have $\grg_{jk}^i\neq 0$. This shows the dependence of
$\grg_{jk}^i$ from the anisotropy terms.

From Euler-Lagrange equations we find for $G^l$ (relation \eqref{Gl}):
\begin{equation} \label{Glreal}
G^l= \frac{1}{2} \connsecond{a}{i}{j}{l} y^{i} y^{j}+
\grs a^{ml} y^{j}\opstwo{j}{m}{A}( \p_{j}\grf(x)\, \unitdn{m})
\end{equation}

Using relation \eqref{Naj} we calculate the nonlinear connection coefficients:
\begin{equation} \label{Najreal}
N^{l}_k=\connsecond{a}{i}{k}{l} y^{i}+
\grs a^{ml}\opstwo{k}{m}{A}\, (\p_k \grf(x)\, \unitdn{m})+
\frac{1}{\grs} a^{ml} y^j \opstwo{j}{m}{A}(\p_j \grf(x)\,
\unitdn{m})  y_k
\end{equation}
or
\begin{equation} \label{Najalt}
N^{l}_k=\overset{(a)}{N} {}^l_j+
\grs a^{ml}\opstwo{k}{m}{A}\, (\p_k \grf(x)\, \unitdn{m})+
\frac{1}{\grs} a^{ml} y^j \opstwo{j}{m}{A}(\p_j \grf(x)\,
\unitdn{m})  y_k
\end{equation}
Relation \eqref{Najalt} clearly shows that the deviation from the Riemann non-linear 
connection is due to anisotropic terms. In the case of an irrotational anisotropic 
field, $\opstwo{k}{m}{A}\, (\p_k \grf(x)\, \unitdn{m})=0$, the non-linear connection is 
identical with the Riemannian one.

The connection coefficients $C^{l}_{ij}$ are given by
\eqref{connectionC}:
\begin{multline} \label{Cijlup}
C^{l}_{ij}= \frac{\grf}{2F}a_{ij}\unitup{l}+ \frac{\grf}{F}
\opstwo{i}{j}{S}(\unitup{i} \grd^{l}_j)- \frac{\grb \grf}{F\grs^2}
\opstwo{i}{j}{S} (\grd^{l}_i y_j)
-\frac{\grf(\grb+m\grs\grf)}{2F^2\grs}a_{ij}y^l-
\\
- \frac{\grf}{2F\grs^2} y_i y_j \unitup{l}-
\frac{\grf(\grs-\grb\grf)}{F^2\grs^2}y^l
\opstwo{i}{j}{S}(\unitdn{i}y_j)-
\left( \frac{\grf}{F}\right)^2
\unitdn{i}\unitdn{j}y^l+ \\
+ \frac{\grf(3\grb+m\grs\grf)}{2F^2\grs^3}
y_i y_j y^l
\end{multline}
Correspondingly, using \eqref{connectionL} we get:

\begin{align}\notag
L^{i}_{jk}&=
\connsecond{a}{i}{j}{l}+\left(
\frac{\grf(\grb+m\grs\grf)}{\grs F^2} y^{a}y^l- \frac{2\grf}{F}
\, \opstwo{a}{l}{S} (y^{a}\unitup{l})
\right)\connfirst{a}{i}{j}{a}+
\frac{\grs}{F} \bigl(
\Lambda_{ij}^{l}+ \\ \notag
&+M_{ij}^{l}
\bigr)
+(
\Lambda_{ija}+M_{ija})\left(
\frac{\grf(\grb+m\grs\grf)}{F^3}y^{a}y^l-
\frac{2\grs\grf}{F^2}
\opstwo{a}{l}{S}(y^{a}\unitup{l})
\right)-\\
&-\left(N^{l}_jC^{i}_{kl}+N^l_k C^{i}_{jl}-f^{ir}N^l_rC_{jkl}
\right)  \label{Lijkreal}
\end{align}
where  $N^{l}_j$ and $C^{i}_{kl}$ are given explicitly by relations
 \eqref{Najreal}, \eqref{Cijlup}.
The curvature of the non linear connection is \eqref{torsion2}:

\begin{align}  \notag
R^{i}_{jk}&=\overset{(a)}{R}{}^{i}_{ajk}y^{a} + \frac{1}{2\grs}
(\p_ka_{mn}\opstwo{j}{b}{A}(\p_j\grf\unitdn{b})-\p_ja_{mn}
\opstwo{k}{b}{A}(\p_k\grf\unitdn{b})
) y^my^na^{bi}+\\  \notag
&+\grs (\p_ka^{bi} \opstwo{j}{b}{A}\p_j\grf\unitdn{b}-
\p_ja^{bi}\opstwo{k}{b}{A}(\p_k\grf\unitdn{b}))
- \grs a^{bi} \opstwo{j}{k}{A}(\p_{bj}\grf\unitdn{k})+\\  \notag
&+\frac{1}{\grs} a^{bi} y^c \left(\opstwo{c}{b}{A}(\p_{kc}\grf\unitdn{b}
)y_j-\opstwo{c}{b}{A}(\p_{jc}\grf\unitdn{b})y_k\right)+\\  \notag
&+
\left( \frac{2}{\grs}
a^{bi}\opstwo{k}{j}{A}(
\p_ky_j
)- \frac{1}{\grs^3}a^{bi}y^my^n\opstwo{k}{j}{A}(\p_ka_{mn}y_j)
\right)\opstwo{a}{b}{A}(\p_a\grf\unitdn{b})y^{a}\\  \notag
&-\grs \left(
\connsecond{a}{k}{b}{i}a^{bc}\opstwo{c}{j}{A}(\p_c\grf\unitdn{j})+
\connsecond{a}{j}{b}{i}a^{bc}\opstwo{k}{c}{A}(\p_k\grf\unitdn{c})
\right)-\grb\p^{i}\grf\opstwo{j}{k}{A}(\p_j\grf\unitdn{k})-\\  \notag
&- \frac{\grb^2+m\grs^2}{2\grs^2}\p^{i}\grf\,
\opstwo{j}{k}{A}(\p_j\grf
y_k)- \frac{\grb}{\grs}\p^{a}\grf\opstwo{k}{j}{A}(\connsecond{a}{a}{k}{i}
y_j)-\\  \notag
&-\frac{1}{\grs} \opstwo{k}{j}{A}(\connfirst{a}{k}{a}{j})\left(
\grb\p^{i}\grf y^{a}-(\p_b\grf y^{b})y^{a}\unitup{i}
\right)
-\\  \notag
&- \frac{1}{2} (\p_b\grf y^{b})\left[
\p^{i}\grf\opstwo{k}{j}{A}(\unitdn{k}y_j)+\unitup{i}
\opstwo{k}{j}{A}(\p_k\grf y_j)
\right] -
\frac{1}{2}(\p_a\grf\p^a\grf)\unitup{i}\opstwo{j}{k}{A}(\unitdn{j}y_{k}) -\\  \notag
&- (\p_a\grf y^{a})\unitup{i}\opstwo{k}{j}{A}(\p_k\grf\unitdn{j})
- \frac{\grb}{2\grs^{2}}(\p_a\grf y^{a})\left[
\p^{i}\grf\opstwo{k}{j}{A}(\unitdn{k}y_j)+\unitup{i}
\opstwo{k}{j}{A}(\p_k\grf y_j)
\right]-\\ \notag
&- \frac{1}{\grs}(\p_a\grf y^{a})\unitup{b}
\opstwo{j}{k}{A}(\connsecond{a}{b}{j}{i}y_k)-
\frac{1}{\grs}\unitdn{a}y^{b}\p^{i}\grf \opstwo{j}{k}{A}(\connsecond{a}{b}{j}{a}y_k)-\\
&- \frac{1}{\grs}\p_a\grf y^b \unitup{i}\opstwo{k}{j}{A}(\connsecond{a}{b}{k}{a}y_j)
- \frac{1}{2\grs^2}(\p_a\grf y^{a})^2 \unitup{i}\opstwo{j}{k}{A}(y_j
\unitdn{k})  \label{Rijk}
\end{align}
where $\overset{(a)}{R}{}^{i}_{ajk}$ is the Riemannian curvature of the metric $a_{ij}$.

The torsion $P^{i}_{jk}$ is given by \eqref{torsion2}:
\begin{equation} \label{Pijk}
P^{i}_{jk}=\connsecond{a}{j}{k}{i}+ \frac{1}{\grs}a^{mi}\left[
\opstwo{l}{m}{A}(\p_l\grf\,\unitdn{m})y_j+a_{jl}
\opstwo{r}{m}{A}(\p_r\grf\,\unitdn{m})y^{r}
\right]-L^{i}_{kj}
\end{equation}
then
\begin{align} \notag
P_{ijk}&= \frac{F}{\grs}\connfirst{a}{j}{k}{i}-
\frac{\grb F}{2\grs^2} a_{jk}\p_i\grf+\left[
\frac{F+m\grs\grf^2}{2\grs^2}\p_i\grf\,y^{i}
- \frac{\grb\grf^2}{2\grs}\p_i\grf\unitup{i}
\right]a_{jk}\unitdn{i}\\ \notag
&+\grf^2\unitdn{l}\connsecond{a}{j}{k}{l}\unitdn{i}
+(\grf^2\unitdn{l}+\frac{\grf}{\grs}y_l)\connsecond{a}{j}{k}{l}
\unitdn{i}-
\frac{F}{2\grs^2}y_j\unitdn{k}\p_i\grf+\\ \notag
&+\left( \frac{\grs+2\grb\grf+m\grs\grf^2}{2\grs^2}
\right)y_j\p_k\grf\,\unitdn{i}
\\ \notag
&-\left(
\frac{\grf^2}{2\grs}(\p_i\grf \,\unitup{i})+
\frac{\grf}{2\grs^2}(\p_i\grf\,y ^{i})
\right)y_j\unitdn{k}\unitdn{i}
+\left(
\frac{\grf}{\grs}\unitdn{l}- \frac{\grb\grf}{\grs^3}y_l
\right)\connsecond{a}{j}{k}{l}y_i+\\ \notag
&+\left(
\frac{m\grf}{2\grs^2}\p_i\grf\, y^{i}-
\frac{\grb\grf}{2\grs^2}\p_i\grf\,\unitup{i}
\right)a_{jk}y_i+ \frac{(m\grs^2-\grb^2)\grf}{2\grs^4}
y_j\p_k\grf\,y_i+\\ \label{Pijk2}
&+\left(
\frac{\grb\grf}{2\grs^4}\p_i\grf\,y^{i}-
\frac{\grf}{2\grs^2}\p_i\grf\,\unitup{i}
\right)y_j\unitdn{k}y_i-\grg_{kji}-\opg{j}{k}{i}(C_{jkl}N^l_i)-f_{ih}L^h_{kj}
\end{align}
The $h$-covariant derivative of the $C_{ijk}$ coefficients is
\begin{align} \label{cijkcov}
C_{ijk|l}=\grd_l C_{ijk}-L^h_{il}C_{hjk}-L^h_{jl}C_{ihk}
-L^h_{kl}C_{ijh}
\end{align}
From relations  \eqref{torsion2},  \eqref{Pijkldn},  \eqref{cartan}, \eqref{Cijlup}, 
\eqref{Lijkreal}, \eqref{Pijk},  and
\eqref{cijkcov}
we can calculate the $P_{ijkl}$ curvature.

Taking into account relations:
\begin{align} \notag
\grd_l L^{i}_{jk}&=\grd_l f^{ir}\left(
\grg_{jkr}-\opg{j}{k}{r}(C_{jkh}N^h_r)
\right)+ f^{ir}
\Bigl(
\grd_l \grg_{jkr}
-\Big[
(\grd_lN^h_j)C_{rkh}+\\ \notag
&+N^h_j(\grd_lC_{rkh})
+(\grd_lN^h_k)C_{jrh}+N^h_k(\grd_l C_{jrh})-
(\grd_l N^h_r)C_{jkh}-\\\label{deltaLijk}
&-N^h_r(\grd_l C_{jkh})
\Big]
\Bigr)\\
\grd_l\grg_{jkr}&=\left(
\frac{1}{\grs}\grd_l F- \frac{F}{\grs^2}\grd_l\grs
\right)\connsecond{a}{j}{k}{r}+
\frac{F}{\grs}\grd_{l}\connsecond{a}{j}{k}{r}+
\grd_l\Lambda_{jkr}+\grd_l M_{jkr}\\ \notag
\grd_kN^i_j&=
\frac{\p \connsecond{a}{j}{r}{i}}{\p x^k}y^r+
\left(
\frac{1}{2\grs} \frac{\p a_{mn}}{\p x^k} y^my^na^{hi}
+\grs \frac{\p a^{hi}}{\p x^k}
\right)\opstwo{j}{h}{A}(\p_j\grf\,\unitdn{h})+\\ \notag
&+ \grs a^{mi}\left[
\opstwo{j}{m}{A}\left((\p^2_{kj}\grf)\,\unitdn{m}
\right)\right]+ \frac{1}{\grs} a^{mi}\left[\opstwo{r}{m}{A}\left(
(\p^2_{kr}\grf)\,\unitdn{m}
\right)\right]y^ry_j+\\ \notag
&+\left(
\frac{1}{\grs} \frac{\p a^{mi}}{\p x^k}-
\frac{1}{2\grs^3} \frac{\p a_{pn}}{\p x^k} y^py^na^{mi}
\right)\opstwo{r}{m}{A}\left(
\p_r\grf\,\unitdn{m}
\right)y^ry_j-\\
&- \frac{\grb}{2} \p^i\grf\,\opstwo{j}{k}{A}\left(
\p_j\grf\,\unitdn{k}
\right)-\grs a^{hl}\connsecond{a}{j}{h}{i}\opstwo{k}{l}{A}
\left(
\p_k\grf\,\unitdn{l}
\right)+ \frac{m}{4}\p^i\grf\,\p_k\grf\,y_j- \notag
\\
&- \frac{1}{4}(\p_a\grf\,\unitup{a})
\p^i\grf\,\unitdn{k}y_j-
\frac{1}{2}\p_h\grf\unitup{i}\left[
a^{hl}\opstwo{k}{l}{A}(\p_k\grf\,\unitdn{l})y_j+
y^{h}\opstwo{k}{j}{A}(\p_k\grf\,\unitdn{j})
\right]- \notag
\\
&-\left(\frac{\grb}{2\grs}\right)^2\p^i\grf\p_j\grf y_k-
\frac{\grb}{2\grs}
\left[
a_{mj}\connsecond{a}{k}{a}{m}y^a\p^i\grf-\p^a\grf\,
\connsecond{a}{j}{a}{i}y_k
\right]-\notag
\\
&- \frac{\grb}{2\grs^2}\p_a\grf\opstwo{i}{a}{A}(\p^i\grf\,\unitup{a})
y_jy_k+ \frac{\grb}{4\grs^2}
(\p_a\grf y^{a})\p^i\grf\,\unitdn{j}y_k-\notag\\
&- \frac{1}{\grs} y^b\connfirst{a}{k}{b}{a}y_j\opstwo{a}{i}{A}(
\p^a\grf\,\unitup{i})-
\frac{1}{2\grs}y^a\Bigl(
\p_a\grf\,\unitup{b}\connsecond{a}{j}{b}{i}y_k+ \notag
\\
&+
(\p_b\grf\,y^b)a_{jl}\connsecond{a}{k}{a}{l}y^a\unitup{i}
\Bigr)+
\frac{\grb}{4\grs^2}(\p_b\grf\,y^b)\unitup{i}\p_j\grf\,y_k+\notag\\
&+ \frac{m}{4\grs^2}(\p_b\grf\,y^b)\p^i\grf\,y_jy_k-
\connsecond{a}{k}{b}{l}\connsecond{a}{j}{l}{i}y^b-\notag\\
&-\frac{1}{2\grs^2}(\p_b\grf\,y^b)\p_a\grf\,a^{al}\unitup{i}y_k
\opstwo{l}{j}{S}(\unitdn{l}y_j)\label{deltaNij}\\
\grd_l\grb&=-\unitdn{h}N^h_l\label{deltagrb}\\
\grd_l\grs&= \frac{1}{2\grs}\p_l a_{ij}y^iy^j-N^h_l(\frac{1}{\grs}a_{ih}y^i)
\\
\grd_l F &= \frac{1}{2\grs}\p_l a_{ij}y^iy^j+\p_l\grf\,\grb
-N^h_l (\frac{1}{\grs}a_{ih}y^i+\grf(x)\unitdn{h})\label{deltaF}
\end{align}
\begin{align}
\grd_l y_i&=-a_{ih} N^{h}_l\\
\grd_l C_{ijk}&=
\frac{3}{2}\left(
\frac{\grf}{\grs^5}\grd_l\grb+
\frac{\grb}{\grs^5}\p_l\grf-5
\frac{\grb\grf}{\grs^6}\grd_l\grs
\right)y_iy_jy_k- \notag
\\
&-
\frac{3\grb\grf}{2\grs^5}\left(a_{ih}N^{h}_ly_jy_k+
a_{jh}N^{h}_ly_iy_k+a_{kh}N^{h}_ly_iy_j\right)+\notag
\\
&+3 \left(
\frac{1}{\grs}\p_l\grf-\frac{\grf}{\grs^2}\grd_l\grs
\right)\opsthree{i}{j}{k}{S}(a_{ij}\unitdn{k})
+ 3\frac{\grf}{\grs}\p_l[\opsthree{i}{j}{k}{S}(a_{ij}\unitdn{k})]-
\notag\\
&- 3\left(
\frac{1}{\grs^3}\p_l\grf-3 \frac{\grf}{\grs^4}\grd_l\grs
\right)\opsthree{i}{j}{k}{S}(y_iy_j\unitdn{k})-
\frac{3\grf}{\grs^3}\grd_l(\opsthree{i}{j}{k}{S}(y_iy_j\unitdn{k}))-
\notag\\
&-3\left(
\frac{\grf}{\grs^3}\grd_l\grb+
\frac{\grb}{\grs^3}\p_l\grf-3
\frac{\grb\grf}{\grs^4}\grd_l\grs
\right)\opsthree{i}{j}{k}{S}(a_{ij}y_k)-
\frac{3\grb\grf}{\grs^3}\grd_l\left(\opsthree{i}{j}{k}{S}
(a_{ij}y_k)
\right)\\
\grd_l\Lambda_{ijk}&=
 \frac{3}{2}
\left(
\frac{\grf}{\grs^5}\grd_l\grb+
\frac{\grb}{\grs^5}\p_l\grf-5
\frac{\grb\grf}{\grs^6}\grd_l\grs
\right) \opg{i}{j}{k}
(y_iy_j\p_ka_{ab})y^ay^b+\notag\\
&+
\frac{3\grb\grf}{2\grs^5}\grd_l\left[
 \opg{i}{j}{k}
(y_iy_j\p_ka_{ab})y^ay^b\right]-\notag\\
&-
\left(
\frac{1}{\grs^3}\p_l\grf-
3 \frac{\grf}{\grs^4}\grd_l\grs
\right)\opg{i}{j}{k}\left(
(y_i\unitdn{j}+y_j\unitdn{i})\p_k a_{ab}
\right)y^ay^b-\notag\\
&- \frac{\grf}{\grs^3}\grd_l\left[
\opg{i}{j}{k}\left(
(y_i\unitdn{j}+y_j\unitdn{i})\p_k a_{ab}
\right)y^ay^b
\right]-\notag\\
&-\left(
\frac{\grf}{4\grs^3}\grd_l\grb+
\frac{\grb}{4\grs^3}\p_l\grf-3
\frac{\grb\grf}{4\grs^4}\grd_l\grs
\right)\opg{i}{j}{k}\left(
a_{ij}\p_ka_{ab}
\right)y^ay^b-\notag\\
&-\frac{\grf\grb}{4\grs^3}
\grd_l
\left[
\opg{i}{j}{k}
\left(
a_{ij}\p_ka_{ab}
\right)
y^ay^b
\right]\\
\grd_l M_{ijk}&=
\frac{1}{2}\left(
\frac{1}{\grs}\grd_l\grb- \frac{\grb}{\grs^2}\grd_l\grs
\right)\opg{i}{j}{k}\left(
a_{ij}\p_k\grf
\right)+ \frac{\grb}{2\grs}\grd_l\left[
\opg{i}{j}{k}\left(
a_{ij}\p_k\grf
\right)\right]-\notag\\
&- \frac{1}{\grs^2}\grd_l\grs\opg{i}{j}{k}
\left(
(y_i\unitdn{j}+y_j\unitdn{i})
\p_k\grf
\right)+
\frac{1}{\grs}\grd_l\left[
\opg{i}{j}{k}
\left(
(y_i\unitdn{j}+y_j\unitdn{i})
\p_k\grf
\right)
\right]-\notag\\
&-
\left(
\frac{1}{\grs^3}\grd_l\grb-3 \frac{\grb}{\grs^4}\grd_l\grs
\right)\opg{i}{j}{k}\left(
y_iy_j\p_k\grf
\right)-
\frac{\grb}{\grs^3}\grd_l\left[
\opg{i}{j}{k}\left(
y_iy_j\p_k\grf
\right)
\right]+\notag\\
&+2\p_l\grf\opg{i}{j}{k}\left(
\unitdn{i}\unitdn{j}\p_k\grf
\right)+2\grf\p_l\left[
\opg{i}{j}{k}\left(
\unitdn{i}\unitdn{j}\p_k\grf
\right)
\right]
\end{align}
and \eqref{Rijkl}, \eqref{Lijkreal}, \eqref{Cijlup}
we may calculate the $R^i_{jkl}$ curvature explicitly from \eqref{Rijk}.\\

The S-curvature \eqref{Sijkldn} is:
\begin{align} \notag
S_{jikh}&= \frac{\grf^2(m\grs^2-\grb^2)}{2F\grs^3}\opstwo{j}{i}{A}
(a_{hj}a_{ik})+ \frac{\grf^2}{2F\grs}\left(
(\opstwo{i}{j}{A}(a_{ki}\unitdn{j})\unitdn{h}+
\opstwo{j}{i}{A}(a_{hj}\unitdn{i})\unitdn{k})
\right)+\\ \notag
&+\frac{\grb\grf^2}{2F\grs^3}\left(
(\opstwo{j}{i}{A}(a_{kj}\unitdn{i})y_h+
\opstwo{k}{h}{A}(a_{jk}\unitdn{h})y_i)
\right)+ \frac{\grf^2}{2F\grs^3}\biggl(
\unitdn{h}y_k\,\opstwo{i}{j}{A}(\unitdn{i}y_j)+\\ \notag
&+ \unitdn{k}y_h
\opstwo{j}{i}{A}(\unitdn{j}y_i)
\biggr)+
\frac{\grb\grf^2}{2F\grs^3}
\left(
\opstwo{h}{k}{A}(a_{ih}\unitdn{k})y_j+
\opstwo{i}{j}{A}(a_{hi}\unitdn{j})y_k\right)+ \\
&+\frac{\grf^2(m\grs^2-2\grb^2)}{4F\grs^5}\biggl(
\opstwo{h}{k}{A}(a_{ih}y_k)y_j+
\opstwo{k}{h}{A}(a_{jk}y_h)y_i \label{Sijkhreal}
\biggr)
\end{align}
\begin{align}
S^r_{ikh}&= \frac{(m\grs^2-\grb^2)}{2F^2\grs^2}
\opstwo{h}{k}{A}(\grd^r_h a_{ki})+
\frac{\grf^2}{2F^2} \left(
\unitdn{i}\opstwo{h}{k}{A}(\grd^r_h\unitdn{k})
+\unitup{r} \opstwo{k}{h}{A}(a_{ki}\unitdn{h})
\right)+ \notag \\ \notag
&+ \frac{\grb\grf^2}{2F^2\grs^2} \left(
\grd^r_k\opstwo{i}{h}{S}(\unitdn{i}y_h)-
\grd^r_h\opstwo{i}{k}{S}(\unitdn{i}y_k)-
\right)+ \frac{\grb\grf^2}{2F^2\grs^2}\unitup{r}\opstwo{h}{k}{A}(a_{ih}y_k)+\\
&+ \frac{(m\grs^2-2\grb^2)\grf^2}{2F^2\grs^4}y_i\opstwo{k}{h}{A}(
\grd^r_ky_h)+ \frac{\grf^2}{2F^2\grs^2}
\biggl(
\unitup{r}y_i\opstwo{k}{h}{A}(\unitdn{k}y_h)+y^r\unitdn{i}
\opstwo{h}{k}{A}(\unitdn{h}y_k)
\biggr)+ \notag \\ \notag
&+ \frac{\grf^2(\grb\grs-\grb^2\grf+2m\grs^2\grf)}{2F^3\grs^2}
y^r\opstwo{h}{k}{A}(a_{ih}\unitdn{k})+\\ \notag
&+
\frac{2\grb^2\grf^2-m\grs^2\grf^2+\grb m \grs\grf^3}{2F^3\grs^3}
y^r\opstwo{k}{h}{A}(a_{ik}y_h)+\\ \label{Supijklreal}
&+ \frac{(m\grs^2-\grb^2)\grf^3}{F^3\grs^4}y^ry_i\opstwo{k}{h}{A}(
\unitdn{k}y_h)
\end{align}
\begin{align} \notag
S_{ih}&= - \frac{3(m\grs^2\grf^2-\grb^2\grf^2)}{4F^2\grs^2}a_{ih}-
\frac{\grf^2}{4F^2}\unitdn{i}\unitdn{h}+\\ \label{Sihreal}
&\qquad\qquad+\frac{\grb\grf^2}{2F^2\grs^2}\opstwo{i}{h}{S}
(\unitdn{i}y_h)+
\frac{3m\grs^2\grf^2-4\grb^2\grf^2}{4F^2\grs^4}y_iy_h\\
 \label{Sreal}
S&=\frac{5(\grb^2-m\grs^2)\grf^2}{2\grs F^3}
\end{align}
From a physical point of view the $S$-curvature can be considered as a curvature 
parameter of anisotropy as it is evident from relation \eqref{Sreal}. In the absence
of anisotropy $\grf=0$, we have $S=0$. In other words, $S$ represents
the measure of anisotropy of matter \cite{Peebles}.

\section{Anisotropic Electromagnetic Field Equations in vacuum}
\label{electro}

The electromagnentic field tensor in special relativity is
$F_{ij}=\p_jA_i(x)-\p_iA_j(x)$.  A generalization in our approach
yields
\begin{align} \notag
\tilde{F}_{ij}&= A_{i|j}(x)- A_{j|i}(x)=
\grd_j A_{i}(x)-\grd_i A_{j}(x)-
L^h_{ij}A_h+ L^h_{ji}A_h\\
&=(\p_j-N^l_j\dot{\p}_l)A_i(x)-
(\p_i-N^l_i\dot{\p}_l)A_j(x) \label{fmn}
\end{align}
or
\begin{equation}
\tilde{F}_{ij}=\p_j A_i(x)-\p_i A_j(x)=F_{ij}
\end{equation}
since $\dot{\p}_lA_i(x)=0$. Therefore the electromagnetic field
tensor remains invariant as in the usual electromagnetic theory of a
Riemannian space-time.

 The first pair of Maxwell equations is
\begin{equation} \label{maxwell1}
\p_lF_{ik}+\p_k F_{li}+\p_i F_{kl}=0
\end{equation}
A generalization of the partial derivatives in our case is to replace them with the
 \mbox{$h$-covariant} derivative of the bundle:
\begin{equation} \label{hcovariant}
\p_lF_{ik} \to F_{ik|l}
\end{equation}
We have
\begin{align} \label{maxwell1seta}
F_{ik|l}&=\grd_lF_{ik}-L^{h}_{li}F_{hk}-L^h_{lk}F_{ih}\\
\label{maxwell1setb}
F_{li|k}&=\grd_kF_{li}-L^{h}_{kl}F_{hi}-L^h_{ki}F_{lh}\\
\label{maxwell1setc}
F_{kl|i}&=\grd_iF_{kl}-L^{h}_{ik}F_{hl}-L^h_{il}F_{kh}
\end{align}
using relations
\[
\grd_lF_{ij}=(\p_l-N^{a}_l\dot{\p}_a)F_{ij},\quad
\dot{\p}_lF_{ij}=0
\]
and
summing \eqref{maxwell1seta}, \eqref{maxwell1setb},
\eqref{maxwell1setc}, yields: $L^{i}_{jk}$
\begin{equation} \label{maxwell1final}
F_{ik|l}+F_{li|k}+ F_{kl|i}=\p_lF_{ik}+\p_k F_{li}+\p_i F_{kl}=0
\end{equation}
where we took into account the symmetric properties of $L^{i}_{jk}$
and $F_{ij}=-F_{ji}$. It is seen
that the first pair of Maxwell equations remains unchanged.

The second pair of Maxwell equations in vacuum  is
\begin{equation} \label{maxwell2}
\p_kF^{ik}=0
\end{equation}
As before we consider:
\begin{equation} \label{hcovariant2}
\p_kF^{ik} \to F^{ik}_{|k}
\end{equation}
Then \eqref{maxwell2} gives:
\begin{equation} \label{maxwell2set}
F^{ik}_{|k}=\grd_kF^{ik}+L^{i}_{hk}F^{hk}+L^{k}_{hi}F^{ih}
\end{equation}
From the antisymmetry of $F^{ik}$ and the symmetry of $L^{i}_{jk}$
the second and third terms of  \eqref{maxwell2set} vanish.
Therefore
 \begin{equation} \label{maxwell2final}
F^{ik}_{|k}=\p_kF^{ik}=0
\end{equation}

From relations \eqref{maxwell2final} and \eqref{maxwell1final}
it is evident that the generalization of the electromagnetic field
 does not change under the presence of the anisotropic model of gravity. It is expected, 
however,
 in analogy with the case of general relativity, that the wave equation should change 
its form. \\

We denote the generalized D'Alambertian by $\dal_F$. In our approach  the D'Alambertian 
is defined by
\begin{equation} \label{Dalambertian}
\dal_F= \frac{1}{\sqrt{-f}} \left(
\grd_{i}\left(\sqrt{-f}f^{ij}\grd_j\right)\right)
\end{equation}
or, equivalently we have:
\begin{equation} \label{dal2}
\dal_F=\grd_i f^{ij}\grd_j+ f^{ij}\grd_i\grd_j+ \frac{1}{2f}
(\grd_i f)f^{ij}\grd_j
\end{equation}
From relation \eqref{*} it follows that
\begin{equation} \label{66}
\grd_if^{ij}=-\left(
L^{i}_{jk}f^{jk}+L^j_{ik}f^{ik}
\right)
\end{equation}
The following relations hold good:
\begin{align} \label{68}
\grd_{i}f&=\p_if-N_i^l\dot{\p}_lf
&\p_if&= \frac{\p f}{\p f_{ab}} \frac{\p f_{ab}}{\p x^{i}}
&\frac{\p f}{\p f_{ab}}&=ff^{ab}\\
\dot{\p}_lf&= \frac{\p f}{\p f_{ab}} \dot{\p}_l f_{ab} \quad\text{then} 
&\dot{\p}_lf&=2ff^{ab}C_{abl}\\
\grd_i f&= ff^{ab}(\p_if_{ab}-2N^l_{i}C_{abl})
&\p_if_{ab}&=\grg_{aib}+\grg_{bia}\end{align}
\begin{align}
\p_if_{ab}-2N^l_iC_{abl}&=
(\grg_{aib}-N^l_iC_{alb})+(\grg_{bia}-N^l_iC_{bla})
\\
L^{i}_{jk}&=\left(\grg^{i}_{jk}-N^l_k C^{i}_{jl}-N_j^{l}C^{i}_{kl}+
f^{ir}N^l_rC_{jkl}
\right) \label{Lijk2}
\end{align}
From relations \eqref{dal2} and \eqref{68}--\eqref{Lijk2}
we have
\begin{equation} \label{dalfinal}
\dal_F=f^{ij}(\grd_i\grd_j-L^k_{ij}\grd_k)
\end{equation}
For the electromagnetic potential $A_l(x)$ relation \eqref{dalfinal} yields
\begin{equation} \label{dalfinal2}
\dal_F A_l(x)=f^{ij}(\p_i\p_j-L^k_{ij}\p_k)A_l(x)=0
\end{equation}

The Lorentz gauge is in special relativity
\begin{equation} \label{Lorentzgauge}
\p_aA^{a}=0
\end{equation}

In a procedure similar to the one used for the D'Alambertian, we find the generalization 
of Lorentz condition:
\begin{equation} \label{Lorentzgeneralization}
\p_aA^{a}=0 \to \frac{1}{\sqrt{-f}}\left(
\grd_a\sqrt{-f}A^{a}
\right)=0
\end{equation}
Again using relations \eqref{dal2} and \eqref{68}--\eqref{Lijk2}, we obtain
equivalently:
\begin{equation} \label{Lorentzfinal}
\p_aA^{a}+L^{a}_{ab}A^{b}=0
\end{equation}
Relation \eqref{Lorentzfinal} is equivalent to $A^a_{|a}=0$

From relations  \eqref{dalfinal2}, \eqref{Lorentzfinal}
 it is evident that the wave equation, as well as the Lorentz condition, are
 modified in such a way as to include anisotropic terms.
However the transformation rule of the $L^{i}_{jk}$ \eqref{transLijk}
connection yields that   the Lorentz condition, and the D'Alambertian are the same for 
any observer.

 It may be possible that relation  \eqref{dalfinal2} is connected with the observed 
anisotropy of the electromagnetic propagation over cosmological distances 
\cite{Nodland}.

Finally we give the equation of motion of a charged particle, subject to the anisotropic 
geometrical framework we developed and to the electromagnetic field (we consider 
$\grs=1$):
\begin{equation*} \label{geodesic+em1}
mc\left(\frac{d^2 x^{l}}{ds^2}+ \connsecond{a}{i}{j}{l}y^{i}y^j+
 a^{lm}(\p_j\grf\unitdn{m}
-\p_m\grf\unitdn{j})y^j\right)= \frac{q}{c} F^{l}_jy^{j}
\end{equation*}
\begin{equation}
 \label{geodesic+em2}
 \frac{d^2 x^{l}}{ds^2}+ \connsecond{a}{i}{j}{l}y^{i}y^j+
 \left(a^{lm}(\p_j\grf\unitdn{m}
-\p_m\grf\unitdn{j})-\frac{q}{mc^2} F^{l}_j
\right) y^j= 0
\end{equation}

It is interesting to note that equation \eqref{geodesic+em2} is
 produced by a Lagrangian of the form
\begin{equation} \label{lagsb}
\Lag=mc\left(\sqrt{a_{ij}y^{i}y^j}+\grf(x) \unitdn{a} y^{a}\right)+ 
\frac{q}{c}A_{a}y^{a}
\end{equation}
Therefore, one may use the Lagrangian \eqref{lagsb} as a metric function and produce  
the equation of motion of a charged particle, subject to an e.m. field and the 
anisotropic geometrical model, as a geodesic of the space generated by \eqref{lagsb}.

\section{Conclusion}

The observed anisotropy of the microwave cosmic rediation, represented by a vector 
$k_{a}(x)$,  can be incorporated in the framework of Finsler geometry.
The equations of geodesics are generalized in a Finsler anisotropic space-time. The 
calculation of a curvature parameter of anisotropy is performed explicitly by the 
contraction of the $S^i_{jkl}$ curvature. Also, the Maxwell equations are unaffected 
from the passage to the anisotropic geometry. The Lorentz condition, as well as the 
generalized D'Alambertian, are shown to be invariant under coordinate transformations.
In our case, however, the generalized wave equation includes the anisotropic vector
through the $L^i_{jk}$ coefficients and the metric tensor $f_{ij}$.


\begin{thebibliography}{99}
\bibitem{Asanov}G.S. Asanov, \emph{Finsler Geometry, Relativity and Gauge Theories}, D. 
Reidel Publishing Company, Dordrecht, Holland, 1985.
\bibitem{kopz}W. Kopczy\'{n}ski, {\em An Anisotropic Universe With Torsion}, Physics 
Letters, 43A (1973).
\bibitem{Lau}L.D. Landau and E.M. Lifshitz, \emph{The Classical Theory of Fields fourth 
edition}, Butterworth Heinemann,  1975.
\bibitem{Miron2}R. Miron, S. Watanabe, S. Ikeda, \emph{Some Connections on Tangent 
Bundle and Their Applications to the General Relativity}, Tensor, N.S., Vol. 46(1987).
\bibitem{Miron1}R. Miron and M. Anastasiei, \emph{The Geometry of
Lagrange Spaces: Theory and Applications}, Kluwer Academic Publishers.
\bibitem{Gravitation}C.W. Misner, K.S. Thorne, J.A. Wheeler, \emph{Gravitation}, W.H. 
Freeman and Company, 1970.
\bibitem{Nodland}B. Nodland and J.P. Ralston,{\em Indication of
Anisotropy in Electromagnetic Propagation over Cosmological
Distances}, Phys. Rev. Lett. {\bf 78}, 3043 (1997).
\bibitem{Peebles}Peebles, P.J.E., \emph{The Large Scale
Structure of the Universe}, Princeton University Press, Princeton, 1980.
\bibitem{Wald}R.M. Wald, \emph{General Relativity}, The University of Chicago Press, 
1984.
\end{thebibliography}
\end{document}